\documentclass[twocolumn,showpacs,preprintnumbers,amsmath,amssymb,superscriptaddress]{revtex4}

\usepackage{epsf}
\usepackage{graphicx}
\usepackage{sidecap}

\usepackage{color}

\def \beq {\begin{equation}}
\def \eeq {\end{equation}}
\begin{document}

\title{{Gigantic surface life-time of an intrinsic topological insulator}}

%Gigantic enhancement of surface life-time of an intrinsic topological insulator}

%Unusually enhanced photoresponse of  an intrinsic topological insulator

%Ultrafast photoresponse of an intrinsic topological insulator

%Optical control of the topological surface state in bulk insulating topological insulator

%Surface electronic structure of a topological Kondo insulator candidate SmB$_6$

\author{Madhab Neupane$^*$}%\footnote{These authors contributed equally to this work.}}
\affiliation {Joseph Henry Laboratory and Department of Physics,
Princeton University, Princeton, New Jersey 08544, USA}
\affiliation {Condensed Matter and Magnet Science Group, Los Alamos National Laboratory, Los Alamos, NM 87545, USA}

\author{Su-Yang Xu$^*$}
\affiliation {Joseph Henry Laboratory and Department of Physics, Princeton University, Princeton, New Jersey 08544, USA}

\author{Yukiaki Ishida$^*$}
\affiliation {ISSP, University of Tokyo, Kashiwa, Chiba 277-8581, Japan}

%\author{T.~Kondo}
%\affiliation {ISSP, University of Tokyo, Kashiwa, Chiba 277-8581, Japan}

\author{Shuang Jia}\affiliation {Department of Chemistry, Princeton University, Princeton, New Jersey 08544, USA}
\affiliation {International Center for Quantum Materials, Peking University, Beijing 100871, China}

\author{Benjamin M. Fregoso}
\affiliation {Department of Physics, University of California, Berkeley, CA, 94720, USA}

%\affiliation {Joint Quantum Institute and Condensed Matter Theory Center, Department of Physics, University of Maryland, College Park, Maryland 20742-4111, USA}

\author{Chang~Liu}
\affiliation {Joseph Henry Laboratory and Department of Physics,
Princeton University, Princeton, New Jersey 08544, USA}

\author{ Ilya Belopolski}
\affiliation {Joseph Henry Laboratory and Department of Physics, Princeton University, Princeton, New Jersey 08544, USA}

\author{Guang Bian}\affiliation {Joseph Henry Laboratory and Department of Physics, Princeton University, Princeton, New Jersey 08544, USA}

\author{Nasser Alidoust}\affiliation {Joseph Henry Laboratory and Department of Physics, Princeton University, Princeton, New Jersey 08544, USA}

%\author{T.-R. Chang}
%\affiliation{Department of Physics, National Tsing Hua University, Hsinchu 30013, Taiwan}

%\author{H.-T. Jeng}
%\affiliation{Department of Physics, National Tsing Hua University, Hsinchu 30013, Taiwan}
%\affiliation{Institute of Physics, Academia Sinica, Taipei 11529, Taiwan}

\author{Tomasz Durakiewicz}
\affiliation {Condensed Matter and Magnet Science Group, Los Alamos National Laboratory, Los Alamos, NM 87545, USA}

%\author{H.~Lin}
%\affiliation {Department of Physics, Northeastern University, Boston, Massachusetts 02115, USA}

\author{Victor Galitski}
\affiliation {Joint Quantum Institute and Condensed Matter Theory Center, Department of Physics,
University of Maryland, College Park, Maryland 20742-4111, USA}
\affiliation {School of Physics, Monash University, Melbourne, Victoria 3800, Australia}

\author{Shik Shin}
\affiliation {ISSP, University of Tokyo, Kashiwa, Chiba 277-8581, Japan}

\author{Robert J. Cava}
\affiliation {Department of Chemistry, Princeton University,
Princeton, New Jersey 08544, USA}

\author{M. Zahid Hasan}
\affiliation {Joseph Henry Laboratory and Department of Physics,
Princeton University, Princeton, New Jersey 08544, USA}
\affiliation {Princeton Center for Complex Materials, Princeton University, Princeton, New Jersey 08544, USA}

\date{\today}
\begin{abstract}

{The interaction between light and novel two-dimensional electronic states holds promise to realize new fundamental physics and optical devices. Here, we use pump-probe photoemission spectroscopy to study the optically-excited Dirac surface states in the bulk-insulating topological insulator Bi$_2$Te$_2$Se, and reveal optical properties that are in sharp contrast to those of bulk-metallic topological insulators. We observe a gigantic optical life-time exceeding 4 $\mu$s  (1 ${\mu}\textrm{s}=10^{-6}$ $\textrm{s}$) for the surface states in Bi$_2$Te$_2$Se, whereas the life-time in most topological insulators such as Bi$_2$Se$_3$ has been limited to a few picoseconds (1 $\textrm{ps}=10^{-12}$ $\textrm{s}$). Moreover, we discover a surface photovoltage, a shift of the chemical potential of the Dirac surface states, as large as $100$ mV. Our results demonstrate a rare platform to study charge excitation and relaxation in energy and momentum space in a two dimensional system.}
% and the observed long-lived optically-lifted Dirac fermions and photovoltage effect pave the way for an optical field effect.} 

%The observed long-lived optically-lifted Dirac fermions and photovoltage effect pave the way for an optical field effect. This new effect can potentially be utilized in optical applications such as a solar battery or a light-induced \textit{p}-\textit{n} transistor based on intrinsic topological insulators.}

\end{abstract}
\date{\today}
\maketitle

%\textbf{I. Introduction}

Two-dimensional electrodynamics shows many interesting effects which do not arise in three dimensions \cite{Graphene, Moore, Hasan, SCZhang,Hsieh, Hasan2, Xia, device, spin_transistor, Neupane_3,Victor, trans, Tanaka, Gedik, SCZ}. Pump-probe photoemission spectroscopy may allow us to directly access many of these two-dimensional phenomena by studying the electronic band structure of a system in an excited state. However, there are currently few experimental systems available where electrons are well confined to move in a plane. Moreover, many of the best-studied two-dimensional electron systems arise at buried interfaces in semiconductor heterostructures and, consequently, are difficult to access by spectroscopy. Meanwhile, graphene has recently offered us a truly two-dimensional electron system \cite{Graphene}, but the low-energy excitations are at the corners of the Brillouin zone and this introduces many challenges in studying two-dimensional electrodynamics by photoemission spectroscopy. Topological insulators (TIs) give rise to surface states and many known TIs have surface states with low-energy excitations at the center of the Brillouin zone. However, most experimentally studied materials suffer from residual bulk conductivity, which dominates transport phenomena and masks the properties of the surface states \cite{Moore, Hasan, SCZhang,Hsieh, Hasan2, Xia}. 

%Here we use pump-probe photoemission spectroscopy to study Bi$_2$Te$_2$Se, a candidate material for a topological insulator with limited bulk conductivity. We observe gigantic lifetime for excited surface states in this system, providing compelling evidence for power-law charge relaxation, which is unique to two-dimensional electrodynamics, and, at the same time, offering a direct optical signature of low bulk conductivity in a topological insulator. This result holds promise for realizing a number of predicted new phases and phenomena based on the interaction between light and the protected surface states of a TI, including an optically-induced topological insulator, the transient topological magneto-electric effect, a topological polariton phase transition and the spin plasmon effect \cite{Victor, trans, Tanaka, Gedik, SCZ}.

\begin{SCfigure*}
\centering
\includegraphics[width=12.8cm]{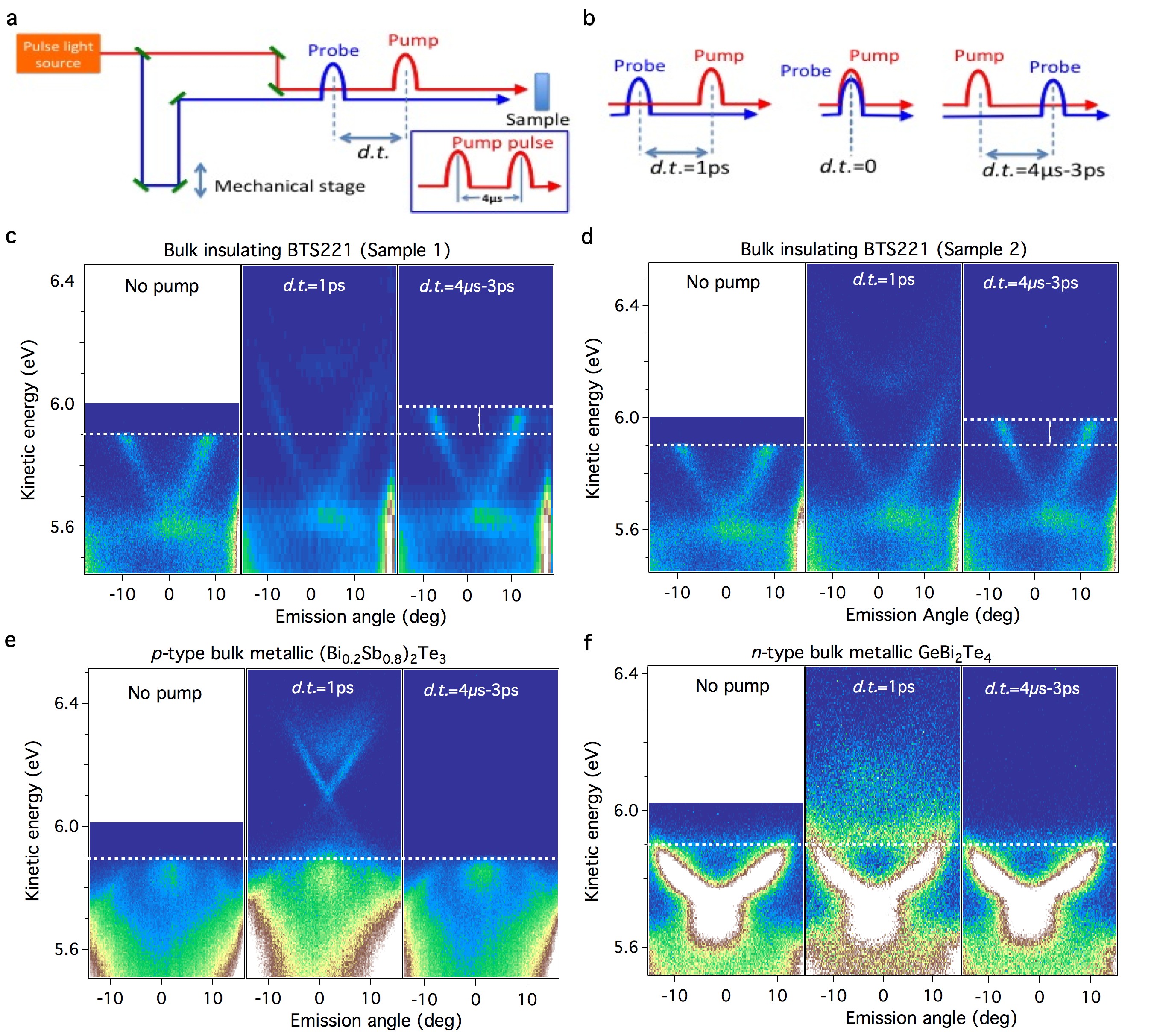}
\caption{{Generation of photo-voltage in a bulk insulating TI}. 
\text{(a,b)} Schematic view of the pump-probe ARPES experimental set up. The delay time is defined in \text{(b)}. The inset in \text{(a)} shows the frequency of the pump pulse. \text{(c)} ARPES band dispersion of bulk insulating Bi$_2$Te$_2$Se (sample 1) near ${\bar\Gamma}$ point measured with no pump (left panel), with pump (middle and right panel).  The delay time is noted on the spectra. \text{(d)} Similar measurement in \text{(c)} for Bi$_2$Te$_2$Se (sample 2).
From \text{(c , d)} the generation of photo-voltage ($\sim$ 100 mV) is evident in the bulk insulating sample. 
Similar measurement as in \text{(d)} for \text{(e)} $p-$type bulk metallic (Bi$_{0.2}$Sb$_{0.8}$)$_2$Te$_3$ TI, and \text{(f)} $n-$ type bulk metallic GeBi$_2$Te$_4$ TI. No generation of the photo-voltage is obtained for the bulk metallic topological insulators (see \text{(e , f)}). The native Fermi level (Fermi level with the absence of pump pulse) is marked by white dash line in the spectra.} 
\end{SCfigure*}

In order to measure the optical life-time of an excited electron, time-resolved angle-resolved spectroscopy (TrARPES) is a powerful tool, where an intense pulse laser (the pump laser) is applied on the surface of a sample to excite electrons from below to above its chemical potential, and then a separate laser beam (the probe laser) is used to study the transient electronic structure as a function of delay time with respect to the pump in a momentum-resolved way. Previously, TrARPES has been used to measure TIs, and a quite short optical life-time of a few picoseconds ($10^{-11}$-$10^{-12}$ s) for the Dirac surface states was reported  \cite{Sobota_1, Hajlaoui, Wang, Crepaldi, Crepaldi_1, Mi, Marsi}. Nevertheless, these experiments were performed with bulk metallic TI samples, which means both the surface states and the bulk conduction bands present at the chemical potential. Therefore, as the pump light is shone, both surface states and the bulk bands were found to be excited and the life-time for both was observed to be as short as $10^{-11}$-$10^{-12}$ s. Such fast dynamics renders the optical control of the surface states a major unsolved challenge in the field. 

%To date, it is not clear how to achieve long-lived excited states in TIs for the theoretically proposed optical applications and novel phenomena.

%These facts make it difficult for any optical control of the surface states. 

%How do we achieve long-lived excited states in TIs for the proposed optical applications?

%To date, it is not clear how to achieve long-lived excited states in TIs for the theoretically proposed optical applications and novel phenomena.
Here, we report a remarkably long life-time exceeding $4$ microseconds (1 ${\mu}$s $=10^{-6}$ s) for the optically excited Dirac surface states. This is achieved by using a truly bulk insulating TI sample Bi$_2$Te$_2$Se (BTS221) with bulk resistivity larger than 6 $\Omega$cm \cite{BTS_Ando, BTS_sdh, Cava_BiTeS, Neupane, Neupane_1, supp}, where only the surface states cross the Fermi level. Surprisingly, we show that the observed long optical life-time in our BTS221 samples leads to a surface photo-voltage (a shift of the chemical potential of the Dirac surface states) as large as $\sim100$ mV, which is also observed to be robust on the time scale of microseconds. 
%A microsecond-time-scale can readily be readout by the modern radio-frequency electronic devices \cite{device_1}, suggesting its potential promise in optical devices based on TIs.
Although surface photo-voltage has been observed in semiconductors \cite{semiconductor, silicon}, we note that our observation of such effect in highly bulk insulating topological insulator is fundamentally different from semiconductor. Topological insulators consist of spin-momentum locked surface states, whereas in semiconductor such topological surface states are completely absent and fundamentally impossible because of the topologically trivial nature of the semiconductor system.

 \begin{SCfigure*}
\centering
\includegraphics[width=13.70cm]{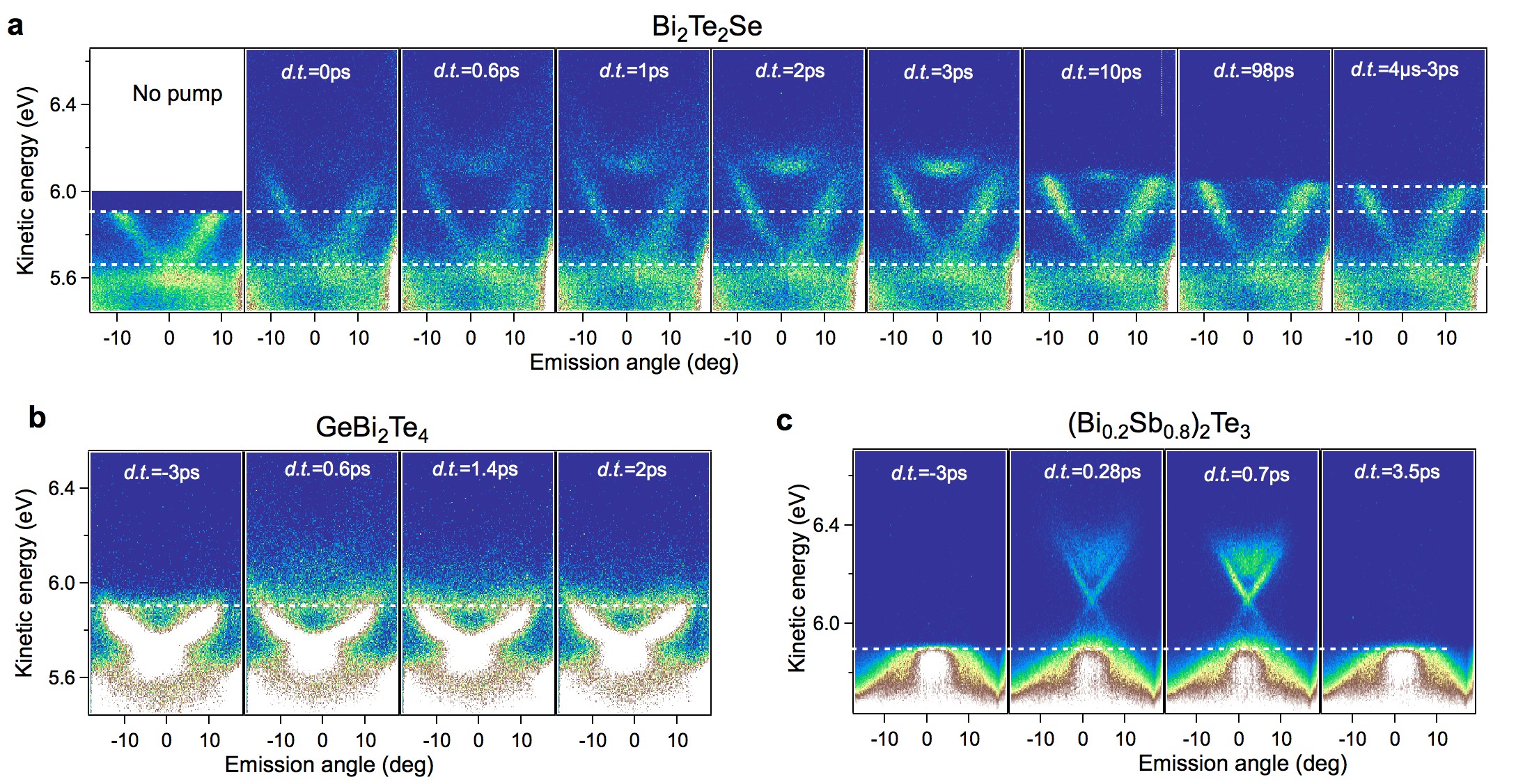}
\caption{{Time-resolved ARPES spectra.} \text{(a)} TrARPES spectra near the ${\bar\Gamma}$ point for various pump-probe delays for BTS221. The populated surface states and bulk conduction band are observed. White dash lines represent the native chemical potential and the kinetic energy position of the Dirac point.
Similar measurement as in \text{(a)} for \text{(b)} GeBi$_2$Te$_4$ and \text{(c)} (Bi$_{0.2}$Sb$_{0.8}$)$_2$Te$_3$. The white dash line in \text{(b)} and \text{(c)} represents the native Fermi level.} 
 \end{SCfigure*}

%\bigskip
%\bigskip
%\textbf{II. Results}

Single crystalline samples of topological insulators used in our measurements were grown using the Bridgman method and characterized by spectroscopic and transport methods, which is detailed elsewhere \cite{BTS_Ando, BTS_sdh, Cava_BiTeS, Neupane, Neupane_1}. The TrARPES setup at the Institute for Solid State Physics (ISSP) in University of Tokyo consists of an amplified Ti:sapphire laser system delivering h$\nu$= 1.47 eV pulses of 170-fs duration with 250-kHz repetition and a hemispherical analyzer \cite{Ishida, supp}. We first describe the basic experimental setup for TrARPES (Figs. 1{(a,b)}) employed in our measurements. Our TrARPES uses a Ti:sapphire laser system, which delivers pulsed light of energy 1.47 eV. Each single pulse is as short as $\sim170$ fs ($1$ fs$=10^{-15}$s), and the pulse is repeated every $4$ $\mu$s (the repetition rate is $250$ KHz). The laser is split into two beams, a pump beam and a probe beam. While the pump laser is applied to the sample directly, the photon energy of probe laser is quadrupled via two $\beta$-BaB$_2$O$_4$ non-linear optical crystals from 1.47 eV to 5.9 eV. The delay time between the pump and the probe laser is tuned by a mechanical stage. Due to a length limit of the mechanical stage, the system can only be probed at a limited range of time delays. Specifically, for our experiment, we can probe the transient electronic structure at the delay time of $0$ to $400$ ps and $4$ $\mu$s$-50$ ps to $4$ $\mu$s. $0$ to $400$ ps means a time period that is immediately after a pump pulse hit the sample. On the other hand, $4$ $\mu$s$-50$ ps  corresponds to a time period that is long after a pump pulse hit the sample, which is immediately before the next pulse. %Details can be found in the Methods Section.

\begin{figure}
\centering
\includegraphics[width=8.50cm]{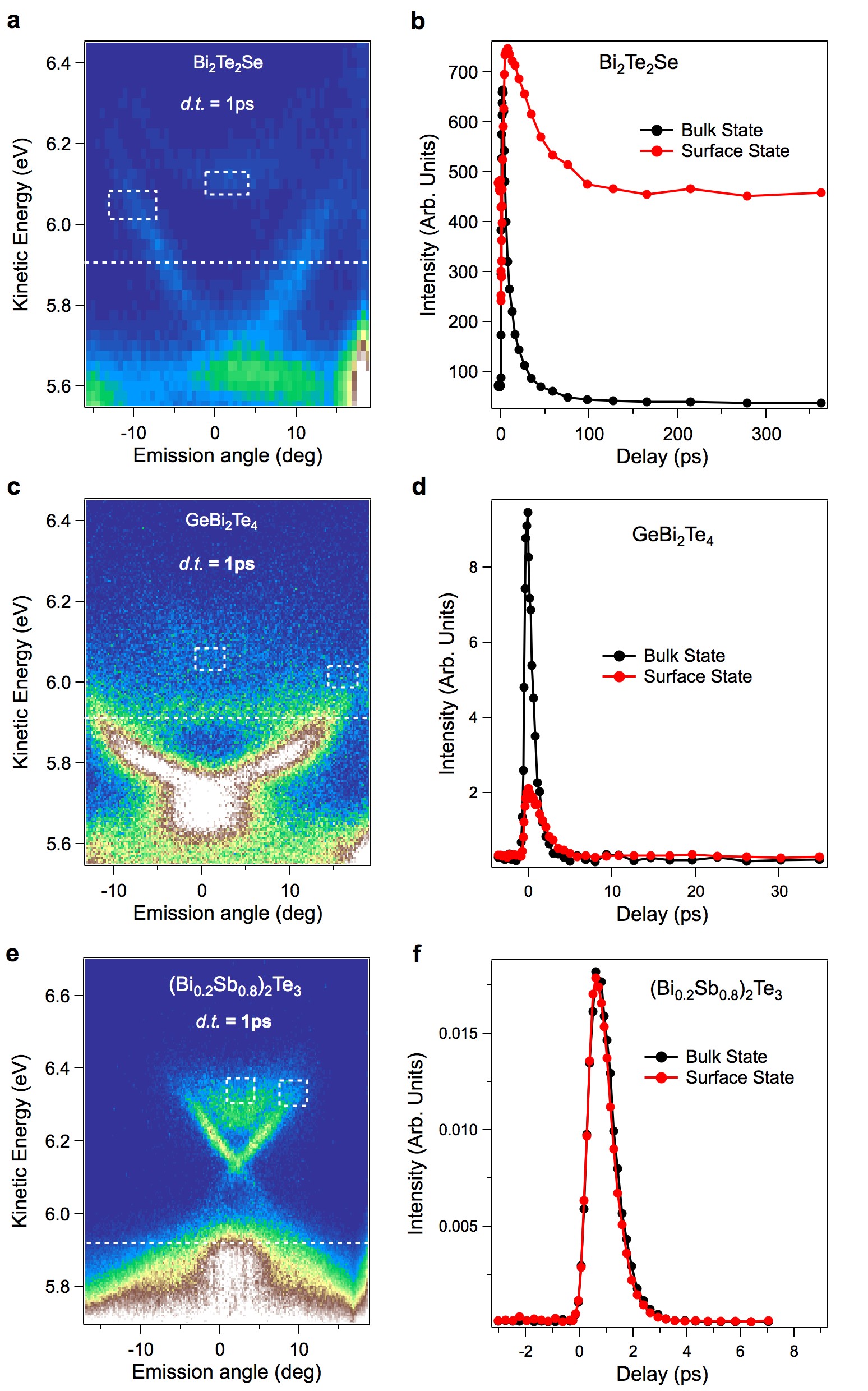}
\caption{{Lifetime comparison between bulk insulating and metallic TIs.} 
\text{(a)} ARPES band dispersion of BTS221 with positive time delay. 
\text{(b)} Ultrafast evolution of the population of surface states (red curve) and bulk states (black curve) for BTS221. Similar measurement as in \text{(a,b)} for  GeBi$_2$Te$_4$ shown in \text{(c,d)} and (Bi$_{0.2}$Sb$_{0.8}$)$_2$Te$_3$  shown in \text{(e,f)}. 
The white rectangles represent the integration window of transient photoemission intensity for surface and bulk states in  \text{(a,c,e)}.}
\end{figure}

%The decay time of SSs of the bulk insulating BTS221 is much longer compared with bulk metallic GBT124 and (Bi$_{0.2}$Sb$_{0.8}$)$_2$Te$_3$.

Figs. 1\text{(c-f)} show ARPES spectra on four different TI samples at representative delay time ($d.t.$) values. Specifically, the four samples are two pieces of BTS221, one (Bi$_{0.2}$Sb$_{0.8}$)$_2$Te$_3$, and one GeBi$_2$Te$_4$. Let us first focus on the ARPES spectra without the pump laser for these four samples. It can be seen from the left panels of Figs. 1\text{(c-f)} that the two BTS221 samples have their chemical potential within the bulk band-gap, and thus only the Dirac surface states are observed to cross the Fermi level, which is consistent with the bulk insulating behavior found in a number of transport experiments \cite{BTS_sdh, Cava_BiTeS, BTS_Ando}. On the other hand, the chemical potential of the (Bi$_{0.2}$Sb$_{0.8}$)$_2$Te$_3$ sample cuts the bulk valence band, which shows its bulk $p$-type metallic character, consistent with previous findings \cite{Hasan,Hasan2}. The GeBi$_2$Te$_4$ sample has chemical potential within the bulk conduction band. Next we consider the transient ARPES spectra at two delay time values ($d.t.=1$ ps and $d.t.=4$ $\mu$s$-3$ ps). For the delay time of $d.t.=1$ ps, which corresponds to the point immediately after the pump laser hits the sample, it can be seen that for all samples, electrons are excited from below to above the chemical potential, and we can observe electronic states above the chemical potential using ARPES (see the middle panels of Figs. 1\text{(c-f)}). Next we study the transient ARPES spectra at $d.t.=4$ $\mu$s$-3$ ps. As shown in the right panels of Figs. 1\text{(c-f)}, for the two metallic samples (the right panels of Figs. 1\text{(e,f)}), the transient ARPES spectra at $d.t.=4$ $\mu$s$-3$ ps is nearly identical with the spectra without the pump within our experimental resolution. This is reasonable because we are measuring the sample nearly $4$ $\mu$s after the pump excitation so we expect that the sample has already returned to the ground state. However, strikingly, for the two bulk insulating BTS samples (the right panels of Figs. 1(c,d), at $d.t.=4$ $\mu$s$-3$ ps the transient ARPES spectra are clearly different from the spectra without pump (the left panels of Figs. 1(c,d) and \cite{supp}). More specifically, we found that the chemical potential of the samples is higher by $\simeq100$ mV. We refer to this shift of the chemical potential due to the pump pulse as a surface photo-voltage. We observe this effect in multiple pieces of BTS221, suggesting that it is not due to systematic errors or artifacts. 

%In order to further exclude any charging effects (it is in fact very unlikely since a charge effect shifts the chemical potential in the negative direction unlike here), in the supplementary information (SI), we show the same measurements as a function of the power of the pump pulse. We observe that the stable surface state photo-voltage of $\simeq100$ mV is independent of the pump power as we vary it from 20 mV to 80 mV.

In order to systematically study how the excited electronic states relax, we present the transient ARPES spectra as a function of delay time ($d.t.$) in Fig. 2. For the two metallic samples (Bi$_{0.2}$Sb$_{0.8}$)$_2$Te$_3$ and GeBi$_2$Te$_4$, as seen in Figs. 2\text{(b,c)}, all excitations relax within $5$ ps. On the other hand, for the bulk insulating BTS221 sample, the relaxation process is longer and more complex. (1) Immediately after the pump ($0{\leq}d.t.{\leq}2$ ps), the electrons are excited to unoccupied states that extend to energies up to $0.8$ eV above the Fermi level. At these short time delays we observe not only the Dirac surface states but also a parabolic bulk conduction band. At these short $d.t.$ values immediately after the pump, the states are quite spread out along the energy axis and it is hard to define a chemical potential \cite{Wang}. (2) For $d.t.{\geq}2$ ps, the excited intensities far from the Fermi level exhibit a very fast decay. Furthermore, a quasi-static chemical potential appears after $2$ ps. And again, strikingly, the quasi-static chemical potential is clearly different from the groundstate chemical potential without the pump, so we observe a surface photo-voltage. (3) As the delay time $d.t.$ increases, the quasi-static chemical potential keeps shifting down. Thus the surface state photo-voltage value decreases. (4) At $d.t.=98$ ps, the bulk conduction band disappears from the spectrum, and a surface photo-voltage of $\sim100$ mV is found. (5) From $d.t.=98$ ps to $d.t.=4$ $\mu$s$-3$ ps, the surface chemical potential only further shifts down by a small energy value ($\sim$ 18 mV) \cite{supp}. Thus the spectrum changes much more quickly for the period $0{\leq}d.t.{\leq}98$ ps as compared to the period of 98 ps${\leq}d.t.{\leq}4$ $\mu$s$-3$ ps. (6) A minor but interesting observation is that the energy position of the bottom of the bulk conduction band is found to move down as a function of $d.t.$ This is consistent with a surface band-bending effect \cite{Analytis2010} as a number of surface state electrons are piled up, which is evident from the observed surface photo-voltage. (7) Moreover, we also find the light-induced change in the surface-state filling; thanks to the detection in the angle-resolved manner with high energy resolution. What we find is not only the long duration of the SPV, but the direct observation of the change in the filling of the surface states that persists for times greater than micro sec time scale. Although the effect of band bending after photoexcitation has been previously discussed in topological insulator materials using TrARPES technique \cite{Marsi, Sobota_1}, the observation of such high surface photo-voltage has not been reported in highly bulk insulating topological insulator material.

%To our knowledge, there is no direct observation of the surface-state dispersions during the SPV generation on topological insulator surface; thanks to TrARPES we were able to monitor the surface band dispersions and their filling. 

%The observation of surface photo-voltage suggests that the light-induced gating is possible in the topological insulator.

In order to quantitatively analyze the rate of change of the transient spectra and obtain the life-time of the excited states, we show the ARPES intensity of the excited states as a function of delay time $d.t.$ (see Fig. 3). To distinguish the life-time between the surface states and the bulk bands, we do this in a momentum-resolved manner. The white boxes in Fig. 3 show the momenta chosen to represent the surface states and the bulk bands, respectively. For the two metallic samples (Bi$_{0.2}$Sb$_{0.8}$)$_2$Te$_3$ and GeBi$_2$Te$_4$, as seen in Figs. 3\text{(c-f)}, the life-time for both the surface states and the bulk bands is found to be below $5$ ps, which is consistent with the life-time reported in previous TrARPES experiments on similar metallic TI samples  \cite{Sobota_1, Hajlaoui, Wang, Crepaldi_1, Mi}. The fact that the life-time for the surface and the bulk is very similar is also intuitive because there are both surface and bulk bands at the Fermi level. As a result, their relaxation processes are coupled. On the other hand, for the bulk insulating BTS221 sample, first of all, the decay is much slower than that of the metallic samples (see Fig. 3\text{(a-b)}). Second, we note that a relatively fast decay (fast with respect to $4$ $\mu$s, but still much slower than the $\sim5$ ps life-time in metallic samples) is found in the first $100$ ps ($0{\leq}d.t.{\leq}98$ ps) for the surface states. After $100$ ps, the surface photo-voltage of $\sim100$ mV decays slowly. In particular, the surface photo-voltage decreases by only $\sim18$ mV from $d.t.=98$ ps to $d.t.=4$ $\mu$s$-3$ ps \cite{supp}. Thus the total life-time for the transient electrons exceeds $4$ $\mu$s. %, although the precise value of the life-time requires further investigation. 

Now we turn to a discussion of  the possible nature of the observed phenomena. First, we note 
that the existence of a surface photo-voltage in our BTS221 samples necessarily implies a charge 
redistribution upon illumination~\cite{Monch2010}. The most likely scenario for this to occur is a 
charge drift caused  by the electric field  of the photo-voltage.
%of a space-charge region near the surface of the TI (see also Supplementary Information). In this scenario, the surface states  (topological and/or trap states) are charged, but since overall charge neutrality must be enforced, the opposite charge spreads over a certain distance away from the surface of the TI into the bulk. This can be related to a band bending near the material's surface. It has been previously observed in Bi$_2$Se$_3$ and can be either upward \cite{Analytis2010} or downward \cite{Bahramy2012}. Our measurements provide evidence of a downward band bending in BTS221. 
We point out some reasons that the excited electrons occupying Dirac surface states in 
bulk insulating samples have exceedingly long lifetimes. 
Upon application of the pump pulse, photo-exited electrons and holes are created.
Their energy gain can be up to the photon energy of $1.47$ eV. 
No electrons escape to the vacuum level because the pump pulse energy 
is lower than the BTS221 work function $\approx 4-5$ eV.
One straight forward relaxation process for the electron hole pairs is the direct recombination across the bulk energy gap. Since our pump pulse is very intense, it generates many electron-hole pairs 
per pulse (the pump power is 20 mW, which means $\sim 10^{17}\times A$ electron-hole pairs per second, where $A$ is the optical absorption coefficient of BTS221), 
and the direct inter-band recombination process  is likely to be inefficient. 
These allows some of the excited electron on BTS221 to drift towards the surface giving rise to a surface photo-voltage. This process is almost complete at about $d.t.$=10 ps as can be seen Fig. 2(a).

For $d.t. >$ 10 ps a quite stable shift in the {surface} chemical potential is observed and 
a slow decay in the ARPES intensity of the excess charge states sets in. This can be understood at qualitative level by a drift 
of charge moving out of the beam spot via  a two-dimensional (2D) relaxation process \cite{Dyakonov1987},
 where excess charge relaxes as  
$\rho(r,t)= (1/2\pi)v/(r^2 + v^2 t^2)^{3/2}$, where $v$ is the drift velocity.
This power-law decay in two dimensions is in sharp contrast to fast exponential decay in three-dimensional metals. 
We estimate the time it takes for the charge to drift out of the pump laser spot (with the length-scale, $D = 1$
mm) under the observed photo-voltage. The drift velocity of the excited electrons is
given by $v=\mu E$, where $\mu$ is the mobility of the Dirac surface states ($\mu\approx 3000$ cm$^2$/Vs, see Ref.~\cite{BTS_sdh}) 
and $E$ is the electric field from the location of the pump beam spot to other locations
on the sample's surface not affected by pumping. Since the quasi-static surface photo-voltage  is about
$0.1$ eV, we set the voltage drop V to be $0.1$ V. So the electric  field $E = V/D = 100$ V/m,
and the drift velocity is $v = \mu E= 30 $ m/s. And finally, the drift time scale is estimated
to be $t=D/v\sim 30$ $\mu$s. To understand the absence of surface photo-voltage in bulk conducting sample GeBi$_2$Te$_4$  and (Bi$_{0.2}$Sb$_{0.8}$)$_2$Te$_3$, we consider the 3D relaxation, which is exponential $\rho(t)=\rho(0)e^{-t/\tau_{3D}}$ where $\tau_{3D}=\epsilon/\sigma$. 
An estimate of $\tau_{3D}$ for our GeBi$_2$Te$_4$  and (Bi$_{0.2}$Sb$_{0.8}$)$_2$Te$_3$ samples (at same density) gives $\tau_{3D}$ in the oder of picosecond which is indeed much shorter than for BTS221 in agreement with our observations( see also \cite{supp} which includes Ref. \cite{Onishi, Madelung1998, Fregoso2013}).

%Finally, we note that a more precise and rigorous estimation of the lifetimes and band bending requires a specialized and delicate theoretical modeling. 

In conclusion, we observe gigantic lifetime for excited surface states in a bulk insulating topological insulator, providing compelling evidence for power-law charge relaxation, which is unique to two-dimensional electrodynamics, and, at the same time, offering a direct optical signature of low bulk conductivity in a topological insulator.  Our observation potentially suggests a
way to develop optical devices, such as an optical p-n
junction (see \cite{supp}) and a solar cell \cite{solar}, based on the optically-excited
Dirac surface states. Our results also provide a paradigm to compare with possible surface photo-voltage 
in other bulk insulating topological insulators  \cite{Dzero, Dai, Neupane_2} as well as to engineer them for future optical devices.

%ensional electrodynamics, and, at the same time, offering a direct optical signature of low bulk conductivity in a topological insulator.  Our observation potentially suggests a way to develop optical devices, based on topological surfaces states, such as an optical p-n junction (see Supplementary Information for the schematic) and a solar cell \cite{solar}, based on the optically-excited Dirac surface states. 

%Regardless of the theoretical origin for our experimental results, our observation of a photo-voltage effect with gigantic lifetime suggests a way to develop optical devices, based on topological surfaces states, such as an optical p-n junction (see Fig. 4(a) inset for the schematic) and a solar cell \cite{solar}, based on the optically-excited Dirac surface states. Our results also provide a paradigm to compare with possible surface photo-voltage  in other bulk insulating topological insulators  \cite{Dzero, Dai, Neupane_2} as well as to engineer them for future optical devices.

\bigskip
\bigskip
\textbf{Acknowledgements}

The work at Princeton and Princeton-led synchrotron X-ray-based measurements are supported by the  Basic Energy Sciences, US Department of Energy (grants DE-FG-02-05ER46200, AC03-76SF00098 and DE-FG02-07ER46352). 
S.S. and Y.I. at ISSP in University of Tokyo acknowledge support from KAKENHI Grants number
23740256 and 2474021. M.N. at LANL acknowledges the support by LANL LDRD program.
B.M.F. acknowledges partial support from Conacyt. V.G. acknowledges support from DOE-BES DESC0001911.
%T.D. at LANL acknowledges support from Department of Energy, Basic Energy Sciences, Division of Material Sciences, and LANL LDRD program. 
T.D. acknowledges support of NSF IR/D program.
M.Z.H. acknowledges Visiting Scientist support from LBNL, Princeton University and the A. P. Sloan Foundation.

\*Correspondence and requests for materials should be addressed to
%\newline
M. N. (Email: mneupane2002@gmail.com) and M.Z.H. (Email: mzhasan@princeton.edu).

\end{document}